\documentclass[aps,prb,twocolumn,amsmath,amssymb,nofootinbib,superscriptaddress,floatfix]{revtex4}

\usepackage[dvips]{color} 
\usepackage{graphicx}
\usepackage{dcolumn} 
\usepackage{bm}      
\usepackage{longtable}

\begin{document}
\title{
Disorder-induced critical phenomena\\
---new universality classes in Anderson localization---  
      }

\author{P.~W.~Brouwer}
\affiliation{Laboratory of Atomic and Solid State Physics, 
             Cornell University, 
             Ithaca, New York 14853-2501, USA}
\author{A.~Furusaki}
\affiliation{Condensed Matter Theory Laboratory, 
             RIKEN, 
             Wako, 
             Saitama 351-0198, 
             Japan}
\author{C.~Mudry}
\affiliation{Paul Scherrer Institute,
             CH-5232 Villigen PSI, 
             Switzerland}
\author{S.~Ryu}
\affiliation{Kavli Institute for Theoretical Physics,
             University of California, Santa Barbara,
             California 93106, USA}
\date{March 8, 2006}

\begin{abstract}
The Anderson metal-insulator transition is a continuous
phase transition driven by disorder.
It remains a challenging problem to theoretically determine
universal critical properties at the transition.
The Anderson transition in a model with a discrete sublattice or
particle-hole symmetry belongs to one of seven universality classes
which are different from the three well-known standard ones.
Here we review our recent theoretical work on these new
universality classes
in (quasi) one and two dimensions.
\end{abstract}

\maketitle

\section{Anderson localization and the Dyson singularity}

According to Landau's Fermi-liquid theory, low-energy electronic
excitations in a solid are weakly interacting quasiparticles 
that are adiabatically connected to excitations in a free electron gas.
By combining Bloch's theorem 
with the Fermi statistics obeyed by electrons, 
one knows if the crystalline state is a perfect metal 
(having an infinite conductivity)
or a band-gap insulator 
at temperatures low enough that the electron gas is degenerate.

It is an empirical fact that there are no perfect metals.
This fact can easily be reconciled with the (nearly) free
electron model by the following arguments due to 
Drude and Sommerfeld. One assumes that the electrons in a metal 
are accelerated by an electric field
for times smaller than the average time $\tau$
needed to scatter off an impurity.
As a result, the distribution of electron velocity reaches a
steady state that is characterized by the finite
Drude-Sommerfeld conductivity
\begin{eqnarray}
\sigma^{\ }_{D}=
\frac{ne^2\tau}{m},
\label{eq: Drude-Sommerfeld theory of conductivity}
\end{eqnarray}
where $e$, $m$, and $n$ are the electron charge, mass, and density,
respectively. This argument is essentially classical.
The assumption underlying
Eq.~(\ref{eq: Drude-Sommerfeld theory of conductivity}) is
that electrons are particles, not waves.
It took two pioneering papers, written by Dyson in 1953%
\cite{Dyson1953}
and Anderson in 1958%
\cite{Anderson1958} 
to call into question this assumption on theoretical grounds.

Anderson's 1958 paper dealt with a quantum particle hopping between
neighboring sites on a three-dimensional lattice. Anderson added
on-site potentials to each lattice site, which were taken to be
random,\cite{Anderson1958} with the standard deviation of the 
random on-site potentials far exceeding the hopping amplitude 
$t$. For this case, Anderson showed that any time-dependent wave
packet that is initially localized both in space and in energy 
almost certainly remains so after an infinite time-evolution. 
Anderson further showed that the wavefunction envelopes decay 
exponentially fast away from their maxima, a phenomenon now
called Anderson localization.
Disorder was thus shown to play a role that 
far exceeds the one that it plays
in the Drude-Sommerfeld picture. 
Disorder alone can turn the nearly free electron gas
(the metallic state) into an insulator
where wavefunctions are localized.
This (Anderson) insulating state is completely 
different in nature from the band-gap insulator.

In the late 1960s, Mott argued that, 
if the randomness of the on-site potential is not too
strong, a mobility edge should separate 
delocalized states with energies close to the band center from 
localized states near the band edges.\cite{Mott1967} 
Upon increasing the disorder strength, 
the two mobility edges approach
each other until they meet at a critical value of the disorder strength. 
The possibility raised by Mott that tuning the Fermi energy through
the mobility edge triggers a transition from the metallic  
to an insulating state solely due to weak disorder in three dimensions
raised the issue of the nature of such a transition.

The decisive argument in support of interpreting
the Anderson metal-insulator transition as a continuous transition
amenable to scaling ideas developed for critical phenomena in
equilibrium phase transitions
came from the scaling theory of Abrahams {\em et al.}\cite{Abrahams1979}
Whereas localization occurs in three dimensions only for strong
enough disorder, 
in one and two-dimensional versions of the Anderson model, eigenfunctions
are always localized, irrespective of the degree of the randomness 
of the on-site potentials.\cite{Hikami1980}
In one dimension, the localization length 
$\xi$ is as small as the mean free path $\ell$. In two
dimensions, $\xi$ is typically much larger than $\ell$. 

Dyson's earlier
1953 paper dealt with a quantum particle hopping along a one-dimensional
lattice (if we use the language of the Anderson model). 
Dyson, however, considered the case that there are no
on-site potentials. Instead, he took the hopping amplitudes
$t_{j,j+1} = t_{j+1,j}$ between neighboring sites $j$
and $j+1$ to be random variables. We refer to Dyson's model 
as the ``random hopping chain''.
The random variations of $t_{j,j+1}$ cause scattering between
Bloch waves, which is
described by means of a mean free time $\tau$. Dyson calculated the
density of states (DOS) per unit length $\rho$ in the thermodynamic
limit and found that it 
diverges near the band center $\varepsilon=0$,
\begin{equation}
  \rho(\varepsilon) = 
  \frac{\rho^{\ }_0(\varepsilon)}
       {|\varepsilon\tau\ln^3|\varepsilon\tau|\,|}
\qquad(|\varepsilon\tau|\ll1).
  \label{eq:dyson}
\end{equation}
Here, $\rho^{\ }_0$ is the DOS in the absence of randomness, which is
finite and nearly constant near $\varepsilon=0$. In contrast,
there is no divergence in the DOS, which is roughly equal
to $\rho_0(\varepsilon)$, in
Anderson's model with random on-site potentials. Further, it 
was found that the conductance $g$ of the random hopping chain 
of length $L$ is not exponentially small 
(as is the case in Anderson's model), but that it has 
an anomalously wide distribution precisely at $\varepsilon=0$, 
with average $\langle g \rangle \sim (\ell/L)^{1/2}$. 
Again, the disorder has far more profoundly changed the 
properties of the chain than one would anticipate based on the
Drude-Sommerfeld model.

The origin of the
difference between the random hopping chain at $\varepsilon=0$ and
Anderson's model with random on-site potentials is a symmetry that is 
present in the former, but absent in the latter. 
In the random hopping chain, 
for each eigenfunction $\psi_j$ with energy $\varepsilon$,
$(-1)^{j}\psi_{j}$ is also an eigenfunction, but with energy
$-\varepsilon$,
and hence the energy eigenvalues always appear in 
pairs, $\pm\varepsilon$.
This symmetry property follows from the fact that
the Hamiltonian of the random hopping chain only has matrix elements 
that connect sites belonging to two distinct sublattices. Under this
``sublattice symmetry'' (SLS), the band center $\varepsilon=0$ is 
special, and wavefunctions at that energy have anomalous localization 
properties. 

The band center of the random hopping chain
can be viewed as an example of two mobility edges merged
together.\cite{Theodorou1976}
Thus, it can serve as a relatively
simple model to study the metal-insulator transition in the Anderson
model in higher dimensions.
The random hopping chain has a diverging
localization length $\xi(\varepsilon) \sim \ell
|\ln |\varepsilon \tau| |$ upon approaching the band
center $\varepsilon=0$.\cite{typical}
In the parlance of critical phenomena, it is convention to call
the exponent $z$ entering the scaling relation $L\sim E^{-1/z}$
between a length scale $L$ and an energy scale $E$ the dynamical scaling
exponent, in which case the scaling relation $\xi(\varepsilon) \sim \ell
|\ln |\varepsilon \tau| |$ corresponds to $z=\infty$
and signals an ``infinite disorder fixed point''.\cite{Fisher1995} 

Whenever the (typical) localization length $\xi$ is much larger than
the mean free path $\ell$, universality is expected. By universality 
is meant that transport and 
thermodynamic characteristics of
a disordered sample only depend on dimensionality and some
intrinsic properties such as the presence or absence of certain
symmetries (an example is the SLS, but also 
time-reversal symmetry (TRS) and spin-rotation symmetry (SRS)
play a role).
Microscopic details such 
as the band parameters and the detailed form of the disorder
should be irrelevant. For example,
the Dyson singularity is universal in that it is independent of the
form of the probability distribution of the hopping randomness (as
long as the standard deviation of the logarithm of the hopping
amplitude is finite). 

The outstanding open question in the problem
of Anderson localization is to compute the evolution of the full probability 
distribution of physical observables as a function of the disorder strength,
and in particular
to compute scaling exponents
at the metal-insulator transition in two and three dimensions.
New insights on disorder induced critical behavior
in the problem of Anderson localization has occurred on two fronts
that we shall review below.
The first front stems from the renewed interest
in the random hopping problem in samples of dimensions larger than
one, and in its three incarnations associated to TRS and SRS.
The second front stems from the introduction by Altland and Zirnbauer 
of four symmetry classes associated to dirty
superconductors.\cite{Altland1997}
In both cases, the existence of disorder induced critical
points was established in a quasi-one dimensional geometry (sample
with a length $L$ much
larger than its width) and shown to be identical
to the Dyson critical point of the random hopping chain.%
\cite{Brouwer1998,Brouwer2000}
Progress also took place in two dimensions with the exact computation of
some critical exponents among these additional symmetry classes of
Anderson localization.%
\cite{Gruzberg1999}
We shall review these developments in the following sections.

\section{Transfer matrix analysis in quasi-one dimension}

Most detailed theoretical results for the localization problem 
have been obtained in one-dimensional and quasi one-dimensional
disordered systems. 
The width of a quasi one-dimensional conductor is defined by
the number of propagating channels $N$ at the Fermi level. 
A one-dimensional conductor has $N=1$. 

The localization problem in (quasi) one dimension has been tackled
using a variety of theoretical methods, most notably field theory, 
network models, and transfer matrix approaches. 
In the field-theoretical approach the localization problem
is mapped onto a field theory known as the (one-dimensional) 
``non-linear sigma model'' (NLSM). 
The NLSM has been the canonical framework to establish universality, 
as the same field theory is obtained for a wide class of microscopic 
disorder.  An essential step in the derivation of one universal
field theory is a separation of length scales (in this case
$\ell \ll \xi =\mathcal{O}(N\ell)$)
which restricts the applicability of this method to quasi one-dimensional
systems with many channels, $N \gg 1$.  The other two methods do not
require the limit $N \gg 1$.  However, equivalence with the NLSM and, 
thus, universality, is to be expected for $N \gg 1$ only.

All three approaches have contributed to the present understanding of 
the localization problem.
The field-theoretical approach suggested the 
existence of critical points induced by disorder belonging to 
different universality classes.\cite{Wegner1976} It has also played an
important role in the characterization of the mesoscopic 
fluctuations.\cite{Mirlin2000}
The mapping onto a network model allowed the numerical computation
of the correlation (localization) length scaling exponent in the 
plateau transition corresponding to the lowest Landau level in the
integer quantum Hall effect (IQHE),\cite{Chalker1988}
and has also been applied to the new universality classes.\cite{Ohtsuki2004}
The transfer matrix approach was implemented numerically to compute the
scaling exponents of the Anderson localization in three dimensions.%
\cite{Kramer1981,Ohtsuki2000}
It also yielded important analytical results for the quasi
one-dimensional case, which we now review.

Let us first illustrate how the transfer matrix approach is applied to
the random hopping problem in one dimension ($N=1$). 
We ignore the electron spin and assume that disorder is weak compared
with the band width.
Since we are interested in properties close to the band center $\varepsilon=0$,
the energy spectrum ($\varepsilon(k)\propto -\cos k$)
can be linearized about it (i.e., about $k=\pm\pi/2$)
and the wavefunction $\psi_j$ can be written as
\begin{equation}
\psi_j\approx
e^{\mathrm{i}\frac{\pi}{2}j}\psi_\mathrm{r}(y)
+e^{-\mathrm{i}\frac{\pi}{2}j}\psi_\mathrm{l}(y)
\label{psi_j}
\end{equation}
with 
$y=j\times$(lattice spacing), and
$\psi^{\ }_{\mathrm{r}}$ and $\psi^{\ }_{\mathrm{l}}$ 
corresponding to waves
moving to the right and to the left, respectively, at velocity $v^{\ }_F$. 
The continuum limit of the Hamiltonian for the random hopping chain
can then be represented by the stationary
Schr\"odinger equation ($\hbar=1$) for a spinor
$\Psi = (\psi^{\ }_{\mathrm{r}}, \psi^{\ }_{\mathrm{l}})^{\rm T}$,
\begin{equation}
\label{eq:schrod}
\left(\mathcal{H}-\varepsilon\right)\Psi=
0,
\quad
\mathcal{H}=
-
\mathrm{i}v^{\ }_{F}\tau^{\ }_{3}\partial^{\ }_{y}
- 
\mathcal{V}(y),
\end{equation}
where $\tau^{\ }_{1,2,3}$ are the Pauli matrices in the
left-mover/right-mover grading.
From Eq.\ (\ref{psi_j}) we infer that multiplying $\psi_j$
by the factor $(-1)^j=e^{\mathrm{i}\pi j}$ amounts to
exchanging $\psi_\mathrm{r}$ and $\psi_\mathrm{l}$, or equivalently,
to the transformation $\Psi\to\tau_1\Psi$.
Hence the SLS of the Hamiltonian, that an eigenfunction $\psi_j$ with energy
$\varepsilon$ has its partner $(-1)^j\psi_j$ with energy $-\varepsilon$,
reads, in the continuum language,
\begin{equation}
  \tau_1 \mathcal{H} \tau_1 = -\mathcal{H},
  \label{eq:SLSH}
\end{equation}
which implies that the scattering potential $\mathcal{V}$ can
have terms proportional to $\tau_2$ and $\tau_3$ only.
Without SLS, $\mathcal{V}$ also has terms 
proportional to $\tau_1$ and 
the $2 \times 2$ unit matrix, which do
not satisfy the symmetry (\ref{eq:SLSH}).

The transfer matrix
$\mathcal{M}_{\varepsilon}(L,0)$ relates the wavefunctions
$\Psi(0)$ and $\Psi(L)$,
\begin{equation}
  \Psi(L) = \mathcal{M}_{\varepsilon}(L,0) \Psi(0).
\end{equation}
Knowledge of $\mathcal{M}$ is sufficient
to calculate the Landauer conductance $g$: the eigenvalues of the 
product $\mathcal{M}^{\ }_\varepsilon \mathcal{M}^{\dagger}_\varepsilon$
come in the pair 
$\exp(\pm 2x)$, which is related to $g$ by
\begin{equation}
  g = \frac{1}{\cosh^2 x}.
  \label{eq:g}
\end{equation}
Transfer matrices satisfy a multiplicative rule:
the transfer matrix of a
disordered segment of length $L_1 + L_2$ is simply the product of
the transfer matrices of the disordered segments of lengths $L_1$
and $L_2$,
\begin{equation}
  \mathcal{M}_\varepsilon(L_1+L_2,0) =
  \mathcal{M}_\varepsilon(L_1+L_2,L_1)
  \mathcal{M}_\varepsilon(L_1,0).
\end{equation}
Using the Born approximation to calculate the
transfer matrix of a segment of length much smaller than the
mean free path $\ell$, the transfer matrix of the full system is then
found by repeated matrix multiplications. 
What makes an analytical solution possible is that each
multiplication changes the transfer matrix only slightly.
As we increase the system size from $L$ to $L+\delta L$,
the random potential $\mathcal{V}(y)$ in the interval
$y\in[L,L+\delta L]$ changes the transfer matrix
from $\mathcal{M}_\varepsilon(L,0)$ into
$\mathcal{M}_\varepsilon(L+\delta L,0)
=\mathcal{M}_\varepsilon(L+\delta L,L)\mathcal{M}_\varepsilon(L,0)$.
Drawing an analogy with the Brownian motion,
we may regard $L$ as a time and the random potential $\mathcal{V}(y)$
as a random force;
$\mathcal{M}^{\ }_\varepsilon(L,0)$ then performs,
as a function of length $L$,
a ``random walk'' in the manifold of allowed
transfer matrices.

The structure of the symmetric space is determined by the
symmetries of the transfer matrix $\mathcal{M}^{\ }_\varepsilon$.
These follow from the fundamental symmetries of the Hamiltonian
$\mathcal{H}$: Hermiticity, TRS, and the SLS. 
For the transfer matrix, these imply
\begin{subequations} 
\begin{eqnarray}
&& 
\mathcal{M}^{\dag}_{\varepsilon}(L)
\tau^{\ }_{3}
\mathcal{M}^{\ }_{\varepsilon}(L)=
\tau^{\ }_{3},
\\
&&
\tau^{\ }_{1}
\mathcal{M}^{* }_{\varepsilon}(L)
\tau^{\ }_{1}=
\mathcal{M}^{\ }_{\varepsilon}(L),
\\
&&
\tau^{\ }_{1}
\mathcal{M}^{\ }_{+\varepsilon}(L)
\tau^{\ }_{1}=
\mathcal{M}^{\ }_{-\varepsilon}(L),
  \label{eq:symmc}
\end{eqnarray}
\end{subequations}
respectively.
Here we have used the shorthand notation
$\mathcal{M}_\varepsilon(L):=\mathcal{M}_\varepsilon(L,0)$.
These conditions define a manifold (Lie group) on which
the transfer matrices $\mathcal{M}_\varepsilon$ live.

\begin{figure}
\begin{center}
\includegraphics[width=9cm,clip]{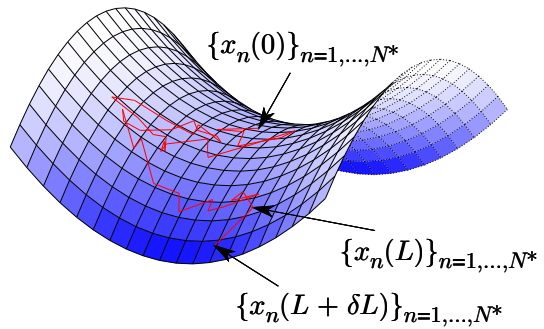}
\caption{
\label{fig: Brown walk on symmetric spaces}
The ``radial coordinate'' of the transfer matrix
makes a Brownian motion on an associated symmetric space.
}
\end{center}
\end{figure}

In view of Eq.\ (\ref{eq:g}), we are 
interested in the radial diffusion only, related to the eigenvalues 
$\exp(\pm 2x)$ of
$\mathcal{M}^{\ }_\varepsilon \mathcal{M}^{\dagger}_\varepsilon$.
Mathematically, this random walk is reformulated
into a random walk on an associated ``symmetric space'',\cite{Helgason}
which is 
obtained after identifying transfer matrices $\mathcal{M}^{\ }_\varepsilon$
that have the same product
$\mathcal{M}^{\ }_\varepsilon \mathcal{M}^{\dagger}_\varepsilon$
(Fig.\ \ref{fig: Brown walk on symmetric spaces}).
Symmetric spaces have a well-defined natural metric,
and if the random potential has 
a Gaussian distribution without long-range correlations,
one finds that the random walk can be described by 
the diffusion equation for 
the symmetric space.\cite{Huffmann}

The SLS (\ref{eq:symmc}) poses an 
extra condition (commutativity with $\tau_1$)
on the transfer matrix at $\varepsilon=0$,
that is absent for transfer matrices at finite
energy or for transfer matrices of a wire without SLS.
In other words, $\varepsilon$ breaks the extra symmetry
of $\mathcal{M}_{\varepsilon=0}$, which we also call SLS.
This fundamentally changes the structure of
the symmetric space and, hence, the solution of the 
localization problem:
at finite energy $\varepsilon$, the radial
coordinate $x$ (\textit{i.e.}, the logarithm of the
eigenvalue of
$\mathcal{M}^{\ }_\varepsilon \mathcal{M}^{\dagger}_\varepsilon$)
performs a {\em biased} random walk, resulting in an exponentially 
small conductance
for large $L$: generically, all wavefunctions are localized 
in one dimension. 
On the other hand, at $\varepsilon=0$ one finds that as $x$ performs an
{\em unbiased} random walk, 
the root-mean-square of the radial coordinate $x$ 
grows proportionally to $\sqrt{L}$ whereas
the typical values of the conductance 
decays as $\exp(-a\sqrt{L/\ell})$
where $a$ is a positive constant.
\cite{Eggarter1978,Fleishman1977} 
Since there is a finite probability
density $\propto L^{-1/2}$ to find $x$ at the origin, the
ensemble averaged conductance decays only algebraically, $\langle
g \rangle \propto L^{-1/2}$.\cite{Stone1981}

\begin{table*} 
\caption{
\label{table 1}
Classification of symmetry classes for
disordered quantum wires. Symmetry classes are defined by the
presence or absence of time-reversal symmetry (TRS) and spin-rotation
symmetry (SRS), and by the further symmetries of the disorder:
sublattice symmetry (SLS) (random hopping model at the band center), and 
particle-hole symmetry (PHS) (zero-energy quasiparticles
in superconductors). For historical reasons, 
the first three rows of the table are referred to
as the orthogonal (O), unitary (U), and symplectic (S) symmetry classes 
when the disorder is generic. The prefix ``ch'' that stands for chiral is
added when the disorder respects a SLS as in the next three rows. Finally,
the last four rows correspond
to dirty superconductors and are named after the symmetric spaces 
associated to their Hamiltonians. 
The table lists the multiplicities of the  
ordinary and long roots $m^{\ }_{o\pm}$ and $m^{\ }_{l}$ of the
symmetric spaces associated with the transfer matrix. Except for
the three chiral classes, one has $m_{o+} = m_{o-} = m_{o}$. For
the chiral classes, one has $m_{o+} = 0$, $m_{o-} = m_{o}$.
The table also lists the degeneracy $D$ of the transfer matrix
eigenvalues, as well as the symbols for the symmetric spaces
associated to the transfer matrix $\mathcal{M}$ and
the Hamiltonian $\mathcal{H}$.
The last three columns list theoretical results for the weak-localization
correction $\delta g$ for $\ell\ll L\ll N\ell$,
the average of $\ln g$ at $L\gg N\ell$, and
the DOS near $\varepsilon=0$. 
The results for $\langle\ln g\rangle$ and $\rho(\varepsilon)$
in the chiral classes refer to the case of $N$ even. For odd $N$,
$\langle\ln g\rangle$ and $\rho(\varepsilon)$ are the same as in
class D. 
       }
\begin{tabular}{lllllllllccc}
\hline 
\hline 
Disorder
&
Class 
& 
TRS
& 
SRS
& 
$m^{\ }_{o}$
& 
$m^{\ }_l$ 
& 
$D$ 
& 
$\mathcal{M}$ 
& 
$\mathcal{H}$
&
$\delta g$
&
$\langle-\ln g\rangle$
&
$\rho(\varepsilon)$ for $0<\varepsilon \tau_c \ll 1$ 
\\ 
\hline
&
O
& 
Yes 
& 
Y 
& 
1 
& 
1 
& 
2 
& 
CI
& 
AI
&
$-2/3$
&
$2L/(\gamma\ell)$
&
$\rho^{\ }_{0}$
\\
generic
& 
U
& 
No 
& 
Y(N)
& 
2 
& 
1 
& 
2(1)
& 
AIII
& 
A 
&
0
&
$2L/(\gamma\ell)$
&
$\rho^{\ }_{0}$
\\
&
S
& 
Y 
& 
N 
& 
4 
& 
1 
& 
2 
& 
DIII
& 
AII
&
$+1/3$
&
$2L/(\gamma\ell)$
&
$\rho^{\ }_{0}$
\\ 
\hline 
&
chO
& 
Y 
& 
Y 
& 
1 
& 
0 
& 
2 
& 
AI
& 
BDI
&
0
&
$2m^{\ }_{o}L/(\gamma\ell)$
& 
$\rho_0 |\ln|\varepsilon\tau_c||$
\\
sublattice
& 
chU
& 
N 
& 
Y(N)
& 
2 
& 
0 
& 
2(1) 
& 
A
& 
AIII
&
0
&
$2m^{\ }_{o}L/(\gamma\ell)$
&
$\pi \rho_0 |\varepsilon \tau_c \ln|\varepsilon\tau_c||$ 
\\
&
chS
& 
Y 
& 
N 
& 
4 
& 
0 
& 
2 
& 
AII
& 
CII
&
0
&
$2m^{\ }_{o}L/(\gamma\ell)$
&
$(\pi \rho_0/3) |(\varepsilon \tau_c)^3\ln|\varepsilon\tau_c||$
\\ 
\hline  
&
CI
& 
Y 
& 
Y 
& 
2 
& 
2 
& 
4 
& 
C
&
CI 
&
$-{4}/{3}$
&
$2m^{\ }_{l}L/(\gamma\ell)$
&
$(\pi\rho_0/2) |\varepsilon \tau_c|$  
\\
particle-hole
& 
C 
& 
N 
& 
Y 
& 
4 
& 
3 
& 
4 
& 
CII
& 
C
&
$-{2}/{3}$
&
$2m^{\ }_{l}L/(\gamma\ell)$
&
$\rho_0 |\varepsilon \tau_c|^2$  
\\
& 
DIII
& 
Y 
& 
N 
& 
2 
& 
0 
& 
2 
& 
D& 
DIII
&
$+{2}/{3}$
&
$4\sqrt{L/(2\pi\gamma\ell)}$
&
$\pi \rho_0/|\varepsilon \tau_c \ln^3|\varepsilon\tau_c||$
\\
& 
D
& 
N 
& 
N 
& 
1 
& 
0 
& 
1 
& 
BDI
& 
D
&
$+{1}/{3}$
&
$4\sqrt{L/(2\pi\gamma\ell)}$
&
$\pi \rho_0/|\varepsilon \tau_c \ln^3|\varepsilon\tau_c||$
\\
\hline
\hline
\end{tabular}
\end{table*} 

A quasi one-dimensional system with spinful electrons is obtained
by replacing the wavefunctions $\psi^{\ }_{\mathrm{r}}$ and 
$\psi^{\ }_{\mathrm{l}}$ by
$2N$-component vectors. The disorder potential $\mathcal{V}$ and the
transfer matrix $\mathcal{M}$ then become $4 N \times 4 N$ matrices.
The eigenvalues of $\mathcal{M}$ come in $D$-degenerate 
pairs $\exp(\pm 2x_n)$, $n=1,2,\ldots,N^*$, where the degeneracy
$D$ depends on the symmetries and $N^* = 2 N/D$. The
Landauer conductance reads
\cite{degeneracy}
\begin{equation}
  g = D \sum_{n=1}^{N^*} \frac{1}{\cosh^2 x_n}.
\label{eq:gD}
\end{equation}
Upon increasing $L$, $\mathcal{M}$ performs a random
walk in the manifold of allowed transfer matrices. If the matrix
elements of the disorder potential have identical and independent
Gaussian distribution without long-range correlations, this random
walk again is described as diffusion on an associated symmetric
space. 
The corresponding diffusion equation (or Fokker-Planck equation)
for the probability
density of the radial coordinates $x_n$, $n=1,\ldots,N^*$ reads
\begin{eqnarray}
\frac{\partial P}{\partial L}=
\frac{1}{2\gamma\ell}
\sum_{j=1}^{N^{*}}
\frac{\partial }{\partial x^{\ }_{j}}
J
\frac{\partial }{\partial x^{\ }_{j}}
J^{-1}
P,
\label{eq: DMPK}
\end{eqnarray}
where $J$ is a Jacobian 
describing the transformation to radial coordinates. It has
the functional form
\begin{equation}
  J = \prod_{j} \sinh^{m_l}(2 x_j) \prod_{k<j} \prod_{\pm}
  \sinh^{m_{o\pm}}(x_j \pm x_k),
\end{equation}
where $m_l$ and $m_{o\pm}$ are numerical constants describing
the geometric structure of the symmetric space (
see Table \ref{table 1}). Finally, $\gamma =
(m_{o+} + m_{o-})(N^*-1)/2 + 1 + m_l$ is another numerical constant.

To determine the appropriate symmetric space, one begins from
the three symmetries listed above (TRS, SRS, and SLS). 
A statistical ensemble of Hamiltonian is classified into symmetry
classes according to its invariance properties under the transformations
induced by these symmetries. 
This gives 6 symmetry classes with the corresponding symmetric
spaces given in the first 6 rows from Table \ref{table 1}. 
In the context of quasi one-dimensional localization for the  standard
symmetry classes (O, U, S in Table \ref{table 1}),
Eq.\ (\ref{eq: DMPK}) was first derived by Dorokhov.\cite{Dorokhov}
It was re-derived independently by Mello, Pereyra, and
Kumar,\cite{MPK}
and is now called the DMPK equation.\cite{Beenakker1997}
The three symmetry classes with SLS are called chiral classes.
Of the three chiral random matrix theories that
describe the energy level statistics on the scale of the mean-level spacing
of the random hopping problem,\cite{Andreev1994}
the chU class has also been applied
to the chiral phase transition in quantum chromodynamics.\cite{QCDreview}
The diffusion equation associated to $\mathcal{M}\mathcal{M}^{\dag}$
for the localization problem with SLS 
was derived by Altland, Simons, and two of the authors.\cite{Brouwer1998}

Four more symmetry classes are obtained\cite{Altland1997} 
when considering the effect of 
disorder on Bogoliubov quasiparticles in a superconductor described 
at the mean-field level.
\cite{self-consistency}
In a superconductor, the quasiparticle wavefunctions at excitation energy
$\varepsilon$ are eigenfunctions 
of the Bogoliubov-de Gennes (BdG) equation,
\begin{equation}
\left(
\mathcal{H}-\varepsilon
\right)\Psi
=0,
\quad
\mathcal{H}=
\left(
\begin{array}{cc}
\hat h & \hat\Delta \\
-\hat\Delta^* & -\hat h^{\mathrm{T}}
\end{array}
\right),
\end{equation}
which has an additional
$2\times2$ grading, the particle-hole grading, when compared to the
Schr\"odinger equation.
Here $\hat h$ ($\hat\Delta$) is a $4N\times4N$ Hermitian (antisymmetric)
matrix representing a single-particle Hamiltonian (superconducting
order parameter).
The Hamiltonian $\mathcal{H}$ 
in the BdG equation satisfies the symmetry relation
$\gamma_1 \mathcal{H}^* \gamma_1 = - \mathcal{H}$ (particle-hole
symmetry), where $\gamma_1$
is the Pauli matrix in the particle-hole grading. For the transfer
matrix particle-hole symmetry implies 
\begin{equation}
  \gamma^{\ }_{1}\mathcal{M}^{\ }_{+\epsilon}(L)\gamma^{\ }_{1}=
  \mathcal{M}^{* }_{-\epsilon}(L).
 \label{eq:MBdG}
\end{equation}
At $\varepsilon=0$, the symmetry requirement
(\ref{eq:MBdG}) modifies the structure of the manifold of allowed
transfer matrices and thus qualitatively modifies the localization
behavior.\cite{Brouwer2000} 
The presence of particle-hole symmetry gives four more
symmetry classes, referred to as C, CI, D, and DIII, depending on 
the presence or absence of TRS and SRS, see Table \ref{table 1}.
At the level
of mean-field theory, charge is not a good quantum number. Only
energy is conserved as is spin in the presence of SRS.
Anderson localization applied to BdG quasiparticles thus aims to
describe the statistics of the global and local DOS of BdG quasiparticles
and the thermal or spin conductance. 
The diffusion equation (\ref{eq: DMPK}) for the four
BdG symmetry classes was reported at the end of the 20th century.
\cite{Brouwer2000}

In total, there are 10 different symmetry classes, yielding
10 symmetric spaces for the transfer matrix ensemble. For four
of those symmetry classes (U, chU, CI, DIII), 
an exact solution of the diffusion
equation (\ref{eq: DMPK}) is possible. For the remaining classes,
asymptotic solutions for $L \ll N\ell$ and $L \gg N\ell$ are known.
We first review the asymptotic for small and large lengths,
and then discuss the exact solutions.

In the regime $\ell \ll L \ll N\ell$ where transport is diffusive,
moments of the conductance $g$ can be obtained
as a power series in $L/N\ell$.
\cite{Beenakker1997,Mudry2000,Imamura2001}
The leading term in the expansion
for the average $\langle g \rangle$ is the Drude conductance
$(\propto N\ell/L)$, 
{\em cf.}\ Eq.~(\ref{eq: Drude-Sommerfeld theory of conductivity}). 
The leading correction to the Drude conductance is known as
the weak localization correction which, unlike
the Drude conductance, depends on symmetry (see Table \ref{table 1}).
When negative, $\delta g$ is
considered as a precursor to exponential localization. 
For the unitary and the three chiral symmetry classes, $\delta g=0$.
In the BdG symmetry classes, 
the sign of $\delta g$ depends on the absence or presence of SRS,
a situation similar to the one for the orthogonal and symplectic cases
in the standard symmetry classes, but the behavior at $L\gg N\ell$
is very different from the standard classes.

The probability distribution of the conductance
becomes very broad if $L \gg N\ell$ and is poorly characterized by
its mean. For the standard symmetry classes, the chiral classes
with an even number of channels, and for the BdG symmetric 
classes CI and C, $\ln g$ becomes
self averaging and thus represents the distribution well. In these
classes, $\ln g$ increases linearly with $L$, corresponding to
exponential localization. We refer
to Table \ref{table 1} for the relevant results. For the 
remaining cases, $ \ln g$ scales proportional to
$(L/N\ell)^{1/2}$ (implying the diverging localization length),
and its fluctuations are of the same order as its mean. 
This is a signature of anomalous localization that reflects
itself with the anomalously slow decay $\sim\sqrt{N\ell/L}$ of the mean
conductance. 

The exact solution of the diffusion equation (\ref{eq: DMPK}) is
possible if the multiplicity of the ordinary root $m_{o}$ is two. 
The exact solution has been used to find the first and second
moments of the conductance for all lengths.
For the chU, CI, and DIII symmetry classes,
the first moment is given by\cite{Brouwer2000,Mudry1999,Lamacraft2004}
\begin{eqnarray}
\langle g\rangle_{\mbox{\begin{tiny}chU\end{tiny}}}&=&
\frac{1}{2s}
+
\frac{1}{s}\sum^\infty_{n=1}
(-1)^{n(N+1)}e^{-\pi^2n^2/8s},
\\
\langle g\rangle_{\mbox{\begin{tiny}CI\end{tiny}}} &=& 
\frac{1}{s}
-
\frac{4}{3}
+ 
4
\sum_{n=1}^{\infty} 
e^{-\pi^2 n^2/4s} 
\left(\frac{1}{s}+\frac{2}{\pi^2 n^2}\right), ~~
\\
\langle g\rangle_{\mbox{\begin{tiny}DIII\end{tiny}}}&=&
\frac{1}{s}
+
\frac{2}{3}
-
4 
\sum_{n=1}^{\infty} 
e^{-\pi^2n^2/2s} 
\frac{1}{\pi^2 n^2},
\end{eqnarray}
respectively, where $s = L/4 N\ell$, in the universal limit 
$N \to \infty$. The exact result shows explicitly that localization is
non-perturbative in $L$, explaining why only very little progress can
be made with standard perturbative techniques.

\begin{figure}
\begin{center}
\includegraphics[width=8cm]{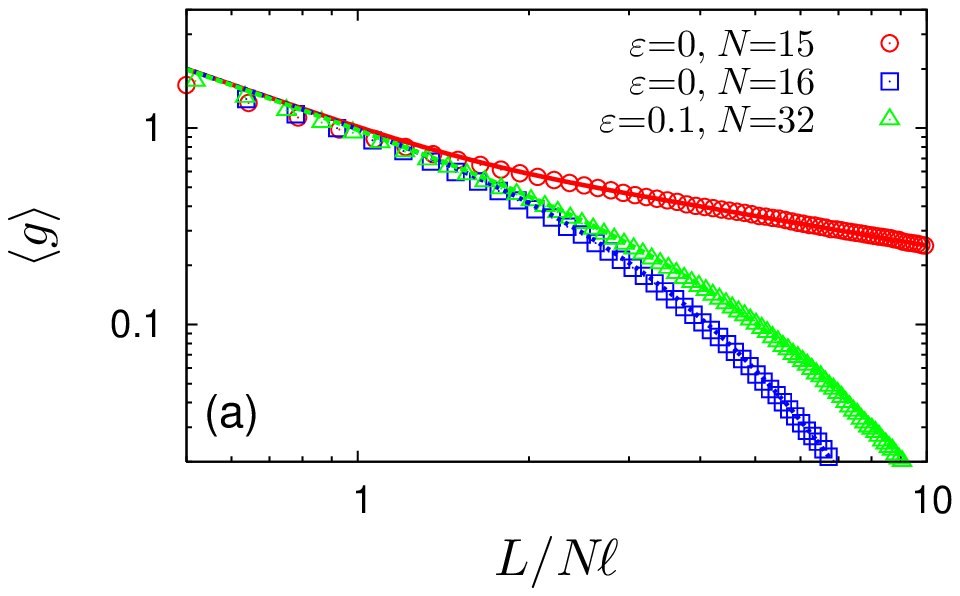}\\
\includegraphics[width=8cm]{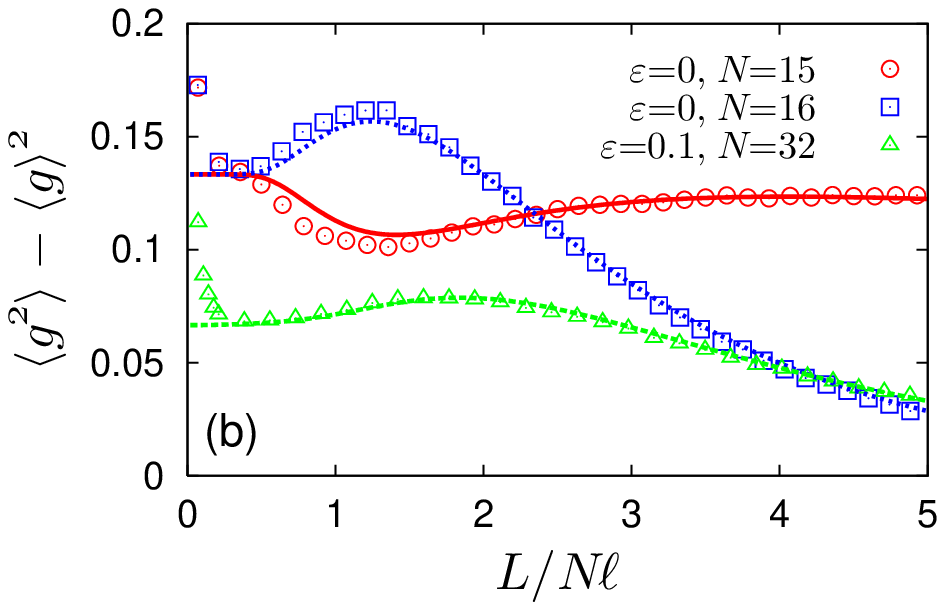}
\caption{
\label{fig: mean and variance of the conductance}
The system size ($L$) dependence of 
(a) mean ($\langle g \rangle$) and (b) variance 
$(\langle g^2\rangle-\langle g\rangle^2)$ of the conductance
in the random flux model.
The numerical results  at the band center $\varepsilon=0$
are denoted by
$\circ$ ($N=15$) and $\square$ ($N=16$),
and the exact solution (at $N\gg1$) from the diffusion equation are 
shown by full 
($N$: odd) and dotted ($N$: even) lines.
For the variance, the result off the band center
$\varepsilon=0.1$
with $N=32$
is also displayed by $\triangle$ for comparison.
It is well-fitted by the analytical result (broken line)
of the unitary symmetric class which is 
obtained by solving the NLSM exactly in quasi-one dimensions.
\cite{Mirlin2000}
}
\end{center}
\end{figure}

In Fig.\ \ref{fig: mean and variance of the conductance}
we show results of numerical simulations of the mean and variance of the
conductance for the so-called quasi-1D random flux model
where $t_{i,j}=e^{\mathrm{i}\theta_{i,j}}$ with random $\theta_{i,j}$
and
which has the SLS but no TRS,
together with the exact results
with $\ell$ as a fitting parameter.
What is specific to the chiral symmetry classes
realized at $\varepsilon=0$
is an even-odd effect in $N$:
as in the random hopping chain,
$\langle g\rangle$ decreases algebraically as $L^{-1/2}$ 
with $L$ for odd $N$
whereas it decays exponentially fast with $L$ for even $N$.
This remarkable even-odd effect is reminiscent
of the one for the spin gap in the number of legs
of $S=1/2$ spin ladder systems.\cite{Dagotto}
For the diffusive regime ($\ell\ll L\ll N\ell$),
the second cumulant of $g$ is called the universal 
conductance fluctuations, and its value in the chiral classes
is twice as large as the one in the corresponding standard symmetry classes.
The numerical results at $\varepsilon\ne0$ show the behavior
of the unitary class as a result of the breaking of
the SLS (\ref{eq:symmc}).

What is the relation between 
the divergence of the localization length and the DOS
singularity originally discovered by Dyson? In order to see
this, one has to adapt the transfer matrix approach to the calculation
of scattering phase shifts. Phase shifts determine the DOS 
through the Friedel sum rule. Interestingly, information about
the scattering phase shifts can be obtained from the standard transfer
matrix approach if we reconsider the transport problem in the 
presence of an imaginary energy $\varepsilon = \mathrm{i}\, \omega$.%
\cite{Titov2001} 
Unlike a real energy, an imaginary energy does not change the symmetry of
the transport problem. The energy $\mathrm{i}\,\omega$ simply yields a
drift term in the diffusion equation (\ref{eq: DMPK}), but otherwise
allows the transport problem to be solved by essentially the same 
methods as outlined previously.
Hence, just like the conductance, the DOS is thus completely determined by 
the geometry of the corresponding symmetric spaces!

For energies of 
the order of
the inverse of the time 
$\tau_c\sim N^2\ell/v^{\ }_F$
needed to
diffuse through a wire of length $\sim N\ell$, the DOS is constant
and given by $\rho^{\ }_{0}= ND/(\pi v^{\ }_{F})$ for all
symmetry classes. This is also the behavior of the DOS for
the standard symmetry classes when $|\varepsilon\tau_{c}|<1$.
For small energies, $|\varepsilon\tau_{c}|< 1$, the remaining symmetry
classes, for which the energy $\varepsilon=0$ plays a special role,
are much more interesting.
The expectation that level repulsion would enforce a vanishing
of the DOS captured by random matrix theory
is wrong whenever the long root $m^{\ }_{l}$
vanishes. 
As shown in table \ref{table 1},
$\rho(\varepsilon)$ shows a power law behavior
with logarithmic corrections whose exponent
for the chiral classes with $N$ even
is not captured by the chiral random matrix theory.
For the chiral classes with $N$ odd and the BdG classes D and DIII,
the DOS exhibits the Dyson singularity.

The diffusion equation (\ref{eq: DMPK}) is a beautiful example of
one-parameter scaling.
Here we'd like to exert one word of caution
about the role played by symmetry: our formal discussion was 
conditional on the assumption that all matrix elements of the weak
potential $\mathcal{V}$ have identical, independent, and Gaussian
distributions. If these conditions are not met, the conductance and
DOS distributions will depend on microscopic details.
Obviously, in such a case, symmetry need not be the only factor in
determining the conductance and DOS distributions.
In particular, if symmetries are only weakly broken (for example,
when TRS is broken by a weak magnetic field
or when a small $\varepsilon>0$ breaks the
SLS of $\mathcal{M}_\varepsilon$\cite{Mudry2000}), one may end up
in a cross-over between two or more of the symmetry classes listed
in Table \ref{table 1}. However, in the limit of large channel
number $N$, universality is believed to be restored. 
Symmetry is 
the only player in the limit $N \gg 1$ (keeping $L/N\ell$ fixed,
to ensure a fixed position in the crossover to localization), so that
the notion of a symmetry class can be elevated to that of a 
universality class.

\section{Progress in two dimensions}

One of the best studied  disorder-induced critical
points in two dimensions
is the plateau transition in the IQHE.
Upon changing the filling fraction of the (lowest) Landau levels
by applying a magnetic field, say,
the localization length for states at the Fermi level
diverges at some critical value $B^{\ }_{c}$,
\begin{eqnarray}
\xi \sim |B-B^{\ }_{c} |^{-\nu},
\end{eqnarray}
at which the Hall conductance jumps by one in the unit of
$e^2/h$.
Here, the exponent for the localization length 
$\nu$ is known numerically to be $2.3<\nu<2.4$ 
within the framework of Anderson localization.\cite{Huckestein1995}
The single-particle wavefunction $\psi(\boldsymbol{r})$ 
at the plateau transition is multifractal, 
\textit{i.e.}, the so-called inverse participation ratio $\mathcal{P}_{q}$
scales with the linear size $L$ of the system as a power law
with the non-linear function of scaling exponents $\tau(q)$,
\begin{eqnarray}
\mathcal{P}_{q}
=
\int_{L^2} \mathrm{d}^{2}
\boldsymbol{r}\,
|\psi(\boldsymbol{r})|^{2q}
\sim
L^{-\tau(q)}.
\end{eqnarray}
The spectrum of exponents
$\tau(q)$ is believed to be universal.

The successful theory
of the plateau transition in the IQHE must predict
$\nu$ and $\tau(q)$. 
Decrypting the critical (conformal field?) theory describing
the plateau transition
remains a tantalizing open problem.
On the other hand, 
there has been some progress on two-dimensional criticality in 
the chiral and BdG universality classes.
In this section, we will review these results with 
main emphasis on exact results.

Within the two-dimensional chiral symmetry class,
the exact computation of the multifractal scaling exponents 
$\tau(q)$ was carried out
for the Hamiltonian describing a particle with relativistic dispersion 
in a random white-noise correlated 
vector potential of vanishing mean
$A_{\mu}(\boldsymbol{r})$,
\begin{eqnarray}
H &=&
\sum_{\mu=1,2}
\sigma^{\ }_{\mu}
\left[
\mathrm{i}\partial^{\ }_{\mu}
+
A^{\ }_{\mu}(\boldsymbol{r})
\right]
\label{eq: def rvp model} 
\end{eqnarray}
($\sigma^{\ }_{\mu}$ is a Pauli matrix).
This problem of Anderson localization is a fine-tuned model 
with symmetry of the chU type that realizes a line of 
critical points as a function of the variance $g_{A}$
of $A_{\mu}$.\cite{Ludwig1994}
The band center $\varepsilon=0$ is a mobility
edge, where a critical state is located, as was the case in the
random hopping chain; see Fig.~3.
The exact computation of the $\tau(q)$-spectrum
for this critical wavefunction is possible 
by exploiting similarities to 
the problem of directed polymers in random media.\cite{Chamon1996}
For $|q| \le \sqrt{2\pi/g_A}$, 
$\tau(q)$ is a quadratic function of $q$ whereas 
it is linear in $q$ outside this region (Fig.~3).
A similar behavior was also observed numerically for 
critical wavefunctions in the IQHE.\cite{Evers2001}
The DOS too can be computed exactly close to the band center.
Its dependence on energy $\varepsilon$ is algebraic,
$\rho(\varepsilon)\sim |\varepsilon|^{\beta}$.
The exponent $\beta$ also displays a non-analytic dependence
on $g^{\ }_{A}$.\cite{Motrunich2002,Horovitz2002}
Both singularities can be ascribed to the non-analyticity
at $g^{\ }_{A}=2\pi$ of the dynamical exponent%
\cite{Horovitz2002,Mudry2003}
\begin{eqnarray}
z(g^{\ }_{A})=
\left\{
\begin{array}{ll}
1
+
\frac{g^{\vphantom{*}}_{A}}{\pi},
&
\hbox{ for }
g^{\vphantom{*}}_{A}<2\pi,
\\
&\\
4\sqrt{\frac{g^{\vphantom{*}}_{A}}{2\pi}}
-
1,
&
\hbox{ for }
g^{\vphantom{*}}_{A}\geq 2\pi.
\end{array}
\right.
\label{eq: dynamical exponent of MDH}
\end{eqnarray}
This non-analyticity as a function of $g^{\vphantom{*}}_{A}$
can be interpreted as a freezing transition
by analogy to the non-analytic dependence on temperature
of the free energy in the random-energy model for spin glasses.\cite{Mudry2003}

\begin{figure}
\begin{tabular}{cc}
\begin{minipage}{0.5\hsize}
\begin{center}
\includegraphics[width=4.5cm,clip]{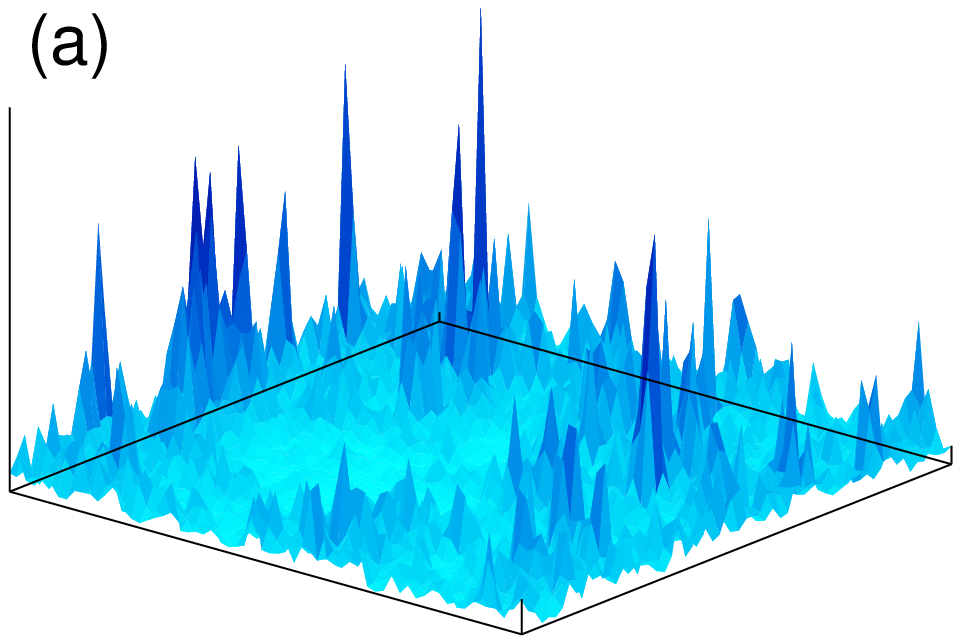}
\end{center}
\end{minipage}
\begin{minipage}{0.5\hsize}
\begin{center}
\includegraphics[width=4.5cm,clip]{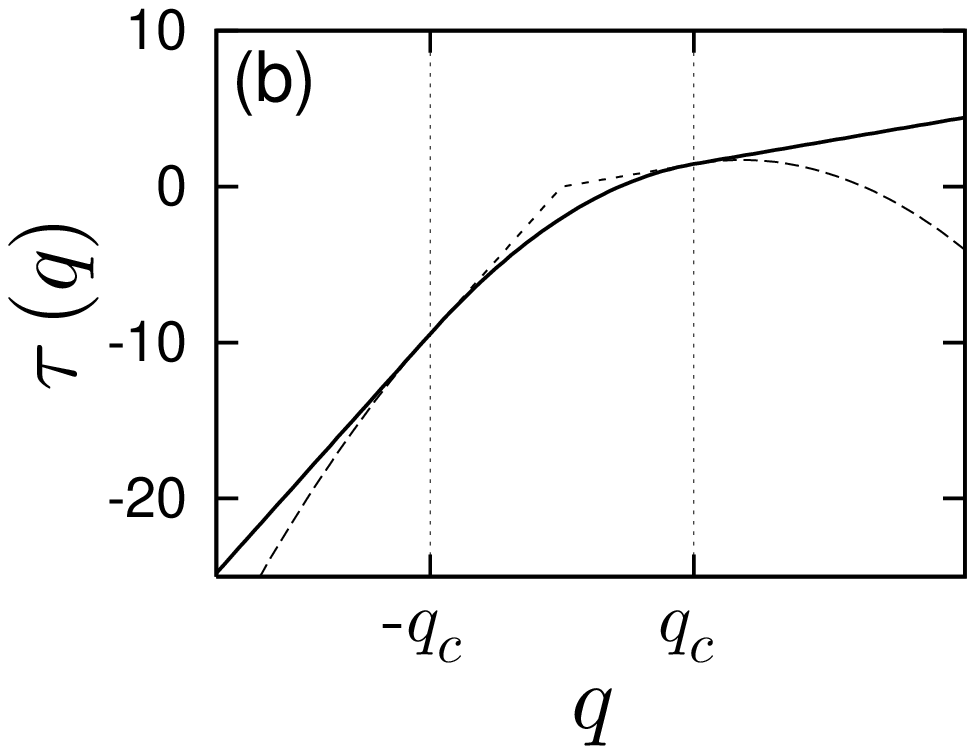}
\end{center}
\end{minipage}
\end{tabular}
\caption{
(a) Typical wavefunction amplitude $|\psi(\boldsymbol{r})|^2$
and (b) its exact $\tau(q)$-spectrum (solid line) for the
random Hamiltonian (\ref{eq: def rvp model}) at $g^{\ }_A=1.2$.
}
\end{figure}

The random vector potential problem (\ref{eq: def rvp model})
is the simplest two-dimensional continuum model with randomness 
that encodes the chiral symmetry. 
For more generic models with SLS, the critical line 
is 
unstable to marginal perturbations (in the renormalization group sense)
compatible with the chiral symmetry,
which renormalize $g^{\ }_A$ to strong coupling.
\cite{Guruswamy2000}
Gade introduced a NLSM 
that describes generic two-dimensional random hopping problems
with SLS, for which she found the diverging DOS
\begin{eqnarray}
\rho(\varepsilon)
\sim
\frac{1}{|\varepsilon|}
\exp(-c |\ln \varepsilon|^{\kappa})
\label{eq: Gade singularity}
\end{eqnarray}
sufficiently close to the band center.
Here, $c$ is a non-universal constant
and 
$\kappa=1/2$
according to Gade.\cite{Gade1993} 
This result can be derived by
dimensional analysis if one assumes that energy and length scales
are related by $\ln\varepsilon\sim-z(L)\ln L$ with
$z(L)\sim\ln L$.
However, Motrunich \textit{et al.}\
conjectured that $\kappa=1/2$ is only a transient as the true asymptotic
dependence of the effective dynamical exponent on $L$ is
$z(L)\sim\sqrt{\ln L}$,
as would follow if the $L$ dependence of $z(L)$
was given by Eq.~(\ref{eq: dynamical exponent of MDH})
with an effective running coupling constant $g^{\ }_{A}\sim\ln L$.%
\cite{Motrunich2002}
If so, the scaling exponent $\kappa=2/3$
follows instead of $\kappa=1/2$.
The latter scaling analysis was confirmed by a functional
renormalization group calculation
for a two-dimensional random hopping problem
in Ref.~\onlinecite{Mudry2003}.

According to the theorem of Anderson on the effect of disorder in
an $s$-wave BCS superconductor, non-magnetic impurities
are largely an irrelevant effect.
In an unconventional BCS superconductor, here defined by an order
parameter with a nonzero angular momentum,
non-magnetic impurities have a much stronger effect. 
With the realizations of unconventional superconductivity
such as high-$T_c$ superconductors,
spin triplet superconductivity in ruthenates,
two-dimensional organic materials,
paired states in the fractional quantum Hall effects
etc., searching for non-perturbative effects in
dirty, unconventional superconductors gained in urgency in the mid 90s.
With SRS and no strong breaking of TRS,
BdG quasiparticles are localized in two dimensions.
In this case the DOS is depressed close to the band center
as expected from level repulsion, 
$\rho(\varepsilon)\sim\varepsilon$ with TRS (CI class) and
$\rho(\varepsilon)\sim\varepsilon^{2}$
without TRS (C class).\cite{Senthil1999}
On the other hand,
as is the case with the IQHE, strong breaking of TRS opens up
the possibility of a new critical point induced by disorder,
the thermal (spin) Hall transition.\cite{Senthil1998}
Using a network model representation of the spin Hall transition
in Class C
Gruzberg \textit{et al.}~reduced the problem of computing the exponent $\nu$
for the diverging localization length and the exponent $\beta$ 
for the power law decay of the DOS to 
the classical problem of bond percolation.
They found $\nu=4/3$ and $\beta=1/7$ at the plateau
transition.\cite{Gruzberg1999}
No results of that magnitude are presently available when SRS
is broken in a dirty BdG but the existence of a plateau transition 
for the Hall thermal current has been established numerically
together with a very rich phase diagram.%
\cite{Chalker2001}

\section{Summary and open problems}

Recent theoretical progress has established that critical
phenomena at mobility edges in quasi-one dimension are
of the Dyson type. Their properties are much better understood than
with Anderson localization in two dimensions in all but few cases.
In this article, we have reviewed two-dimensional models 
for which the multifractal spectrum or some critical indices can be
obtained exactly. Within a classification based on symmetries
of disorder induced critical points in the problem of Anderson localization,
the chiral symmetry class appears to be the simplest both in quasi
one and two dimensions.

The physics of the chiral symmetry classes is closely related to
the physics of the pseudogap in Peierls systems and to the physics
of the chiral phase transition in quantum chromodynamics, 
which we could not discuss in this limited space.
Chiral and BdG classes are also related to classical and quantum
random spin systems. For examples,
the Dyson singularity in the random hopping chain
is connected to the random singlet phase in one-dimensional 
quantum spin systems,\cite{Dagotto}
the two-dimensional chiral symmetry classes are
related to the two-dimensional gauge glass
model,\cite{Guruswamy2000,Carpentier2000}
and the classical random bond Ising model in two dimensions
can be mapped onto a two-dimensional network model belonging to the 
symmetry class D. In the last example, this connection
was used to demonstrate the multifractal behavior\cite{Merz2002} 
of the correlation function of a dual order parameter
and a freezing phenomenon.\cite{Mildenberger2005}

Another important open issue is that of the competition 
between disorder and interactions in the chiral and BdG universality classes.
The situation seems to be under better control for 
the BdG symmetry classes than in the standard symmetry classes.%
\cite{Fabrizio2002}


\begin{thebibliography}{99}

\bibitem{Dyson1953}
F.\ J.\ Dyson:
Phys.\ Rev.\ \textbf{92} (1953) 1331.

\bibitem{Anderson1958}
P.\ W.\ Anderson:
Phys.\ Rev.\ \textbf{109} (1957) 1492.

\bibitem{Mott1967}
N.\ F.\ Mott:
Adv.\ Phys.\ \textbf{16} (1967) 49;
Phil.\ Mag.\ \textbf{17} (1968) 1259.

\bibitem{Abrahams1979}
E.\ Abrahams, P.\ W.\ Anderson, D.\ C.\ Licciardello
and T.\ V.\ Ramakrishnan:
Phys.\ Rev.\ Lett.\ \textbf{42} (1979) 673.

\bibitem{Hikami1980}
Anderson transition can occur even in two dimensions
when strong spin-orbit interaction is present.
S.\ Hikami, A.\ I.\ Larkin and Y.\ Nagaoka:
Prog.\ Theor.\ Phys.\ \textbf{63} (1980) 707.

\bibitem{Theodorou1976}
G.\ Theodorou and M.\ H.\ Cohen:
Phys.\ Rev.\ B \textbf{13} (1976) 4597.

\bibitem{typical}
In order to do justice
to the broadness of the distributions of DOS
and localization length, one distinguishes a ``typical localization 
length'' $\propto |\ln |\varepsilon \tau||$ and an ``average localization 
length'' $\propto \ln^2|\varepsilon \tau|$ :
L.\ Balents and M.\ P.\ A.\ Fisher:
Phys.\ Rev.\ B \textbf{56} (1997) 12970;
S.\ Ryu, C.\ Mudry and A.\ Furusaki:
\prb \textbf{70} (2004) 195329.

\bibitem{Fisher1995}
D.\ S.\ Fisher:
Phys.\ Rev.\ B \textbf{50} (1994) 3799;
Phys.\ Rev.\ B \textbf{51} (1995) 6411.

\bibitem{Altland1997}
A.\ Altland and M.\ R.\ Zirnbauer:
Phys.\ Rev.\ B \textbf{55} (1997) 1142.

\bibitem{Brouwer1998}
P.\ W.\ Brouwer, C.\ Mudry, B.\ D.\ Simons and A.\ Altland:
Phys.\ Rev.\ Lett.\ \textbf{81} (1998) 862.

\bibitem{Brouwer2000}
P.\ W.\ Brouwer, A.\ Furusaki, I.\ A.\ Gruzberg and C.\ Mudry:
Phys.\ Rev.\ Lett.\ \textbf{85} (2000) 1064.

\bibitem{Gruzberg1999}
I.\ A.\ Gruzberg, A.\ W.\ W.\ Ludwig and N.\ Read:
Phys.\ Rev.\ Lett.\ \textbf{82} (1999) 4524.

\bibitem{Wegner1976}
F.\ J.\ Wegner: 
Z.\ Phys.\ B \textbf{25} (1976) 327;
Z.\ Phys.\ B \textbf{35} (1979) 207.

\bibitem{Mirlin2000}
A.\ D.\ Mirlin: 
Phys.\ Rep.\ \textbf{326} (2000) 259.

\bibitem{Chalker1988}
J.\ T.\ Chalker and P.\ D.\ Coddington:
J.\ Phys.\ C \textbf{21} (1988) 2665.

\bibitem{Ohtsuki2004}
B.\ Kramer, T.\ Ohtsuki and S.\ Kettemann:
Phys.\ Rep.\ \textbf{417} (2005) 211.

\bibitem{Kramer1981}
A.\ McKinnon and B.\ Kramer:
Phys.\ Rev.\ Lett.\ \textbf{47} (1981) 1546 .

\bibitem{Ohtsuki2000}
See for example,
T.\ Ohtsuki,
\textit{Metal-insulator transitions in disordered electron systems}
in 
\textit{Trends in Modern Physics 2}
(in Japanese),
(Kyoritsu, 2000).

\bibitem{Helgason}
S.\ Helgason: \textit{Differential geometry, Lie groups and
symmetric spaces} (Academic Press, New York, 1978).

\bibitem{Huffmann}
A.~H\"uffmann:
J.\ Phys.\ A \textbf{23} (1990) 5733.


\bibitem{Eggarter1978}
T.\ P.\ Eggarter and R.\ Riedinger:
Phys.\ Rev.\ \textbf{18} (1978) 569.

\bibitem{Fleishman1977}
L.\ Fleishman and D.\ C.\ Licciardello:
J.\ Phys.\ C \textbf{10} (1977) L125.

\bibitem{Stone1981}
A.\ D.\ Stone and J.\ D.\ Joannopoulos:
Phys.\ Rev.\ B \textbf{24} (1981) 3592.

\bibitem{degeneracy}
For standard and chiral symmetry classes,
it is customary not to include the degeneracy $D$
in the definition of the Landauer formula Eq.\ (\ref{eq:gD}).
Reader should be careful when comparing the results to the
literature.

\bibitem{Dorokhov}
O.\ N.\ Dorokhov:
JETP Letters \textbf{36} (1982) 318.

\bibitem{MPK} 
P.\ A.\ Mello, P.\ Pereyra and N.\ Kumar:
Ann.\ Phys.\ (NY) \textbf{181} (1988) 290.

\bibitem{Beenakker1997}
C.\ W.\ J.\ Beenakker: Rev.\ Mod.\ Phys.\ \textbf{69}
(1997) 731.

\bibitem{Andreev1994}
K.\ Slevin and T.\ Nagao: Phys.\ Rev.\ Lett.\ \textbf{70} (1993) 635;
A.\ V.\ Andreev, B.\ D.\ Simons and N.\ Taniguchi:
Nucl.\ Phys.\ B \textbf{432} (1994) 487.

\bibitem{QCDreview}
See for example,
J.\ J.\ M.\ Verbaarschot: \texttt{hep-th/0502029.}

\bibitem{self-consistency}
We do not impose the condition of self-consistency on the 
superconducting gap here.

\bibitem{Mudry2000}
C.\ Mudry, P.\ W.\ Brouwer and A.\ Furusaki:
Phys.\ Rev.\ B \textbf{62} (2000) 8249.

\bibitem{Imamura2001}
T.\ Imamura and K.\ Hikami:
J.\ Phys.\ Soc.\ Jpn.\ \textbf{70} (2001) 3312.

\bibitem{Mudry1999}
C.\ Mudry, P.\ W.\ Brouwer and A.\ Furusaki:
Phys.\ Rev.\ B \textbf{59} (1999) 13221.

\bibitem{Lamacraft2004}
A.\ Lamacraft, B.\ D.\ Simons and M.\ R.\ Zirnbauer:
Phys.\ Rev.\ B \textbf{70} (2004) 075412.

\bibitem{Dagotto}
E.\ Dagotto and T.\ M.\ Rice:
Science \textbf{271} (1996) 618.

\bibitem{Titov2001}
M.\ Titov, P.\ W.\ Brouwer, A.\ Furusaki and C.\ Mudry:
Phys.\ Rev.\ B \textbf{63} (2001) 235318.

\bibitem{Huckestein1995}
B.\ Huckestein:
Rev.\ Mod.\ Phys.\ \textbf{67} (1995) 357.

\bibitem{Ludwig1994}
A.\ W.\ W.\ Ludwig, M.\ P.\ A.\ Fisher, R.\ Shankar and G.\ Grinstein:
Phys.\ Rev.\ B \textbf{50} (1994) 7526.

\bibitem{Chamon1996}
C.\ C.\ Chamon, C.\ Mudry and X.-G.\ Wen:
Phys.\ Rev.\ Lett.\ \textbf{77} (1996) 4194.

\bibitem{Evers2001}
F.\ Evers, A.\ Mildenberger and A.\ D.\ Mirlin:
Phys.\ Rev.\ B \textbf{64} (2001) 241303(R).

\bibitem{Motrunich2002}
O.\ Motrunich, K.\ Damle and D.\ A.\ Huse:
Phys.\ Rev.\ B \textbf{65} (2002) 064206.

\bibitem{Horovitz2002}
B.\ Horovitz and P.\ Le Doussal:
Phys.\ Rev.\ B \textbf{65} (2002) 125323.

\bibitem{Mudry2003}
C.\ Mudry, S.\ Ryu and A.\ Furusaki:
Phys.\ Rev.\ B \textbf{67} (2003) 064202.

\bibitem{Guruswamy2000}
S.\ Guruswamy, A.\ LeClair and A.\ W.\ W.\ Ludwig:
Nucl.\ Phys.\ \textbf{B583} (2000) 475.

\bibitem{Gade1993} 
R.\ Gade: 
Nucl.\ Phys.\ B \textbf{398} (1993) 499.

\bibitem{Senthil1999}
T.\ Senthil and M.\ P.\ A.\ Fisher:
Phys.\ Rev.\ B \textbf{60} (1999) 6893.

\bibitem{Senthil1998}
T.\ Senthil, M.\ P.\ A.\ Fisher, L.\ Balents and C.\ Nayak:
Phys.\ Rev.\ Lett.\ \textbf{81} (1998) 4704.


\bibitem{Chalker2001}
J.\ T.\ Chalker, N.\ Read, V.\ Kagalovsky, 
B.\ Horovitz, Y.\ Avishai and A.\ W.\ W.\ Ludwig:
Phys.\ Rev.\ B \textbf{65} (2001) 012506.

\bibitem{Carpentier2000}
D.\ Carpentier and P.\ Le Doussal:
Nucl.\ Phys.\ B \textbf{588} (2000) 565.

\bibitem{Merz2002}
F.\ Merz and J.\ T.\ Chalker:
Phys.\ Rev.\ B \textbf{65} (2002) 054425.

\bibitem{Mildenberger2005}
A.\ Mildenberger, F.\ Evers, R.\ Narayanan, A.\ D.\ Mirlin
 and K.\ Damle:
Phys.\ Rev.\ B \textbf{73} (2006) 121301(R).

\bibitem{Fabrizio2002}
For example, it was reported that 
the renormalization group flow of the NLSM
is simpler for the BdG classes than for the standard ones:
M.\ Fabrizio, L.\ Dell'Anna and C.\ Castellani:
Phys.\ Rev.\ Lett.\ \textbf{88} (2002) 076603;
M.\ Jeng, A.\ W.\ W.\ Ludwig,
T.\ Senthil and C.\ Chamon:
\texttt{cond-mat/0112044}.

\end{thebibliography}
\end{document}